\documentclass[prd,twocolumn,nofootinbib,showpacs,amsmath,amssymb]{revtex4}
\usepackage{graphicx}
\usepackage{dcolumn}
\usepackage{bm}
\usepackage{latexsym}
\usepackage{amssymb}
\usepackage{amsmath}
\usepackage{subfigure}
\usepackage{hyperref}

\begin{document}
	
	\title{The ratio $\rho^{pp}_{\bar  pp}(s)$ in Froissaron and Maximal Odderon approach}
	
	\author{E. Martynov and G. Tersimonov}
	
	\affiliation{%
		Bogolyubov Institute for Theoretical Physics,\\ 03143, Metrologicheskaja st. 14b, Kiev-143, Ukraine.\\
		e-mails: martynov@bitp.kiev.ua;  gters.hep@gmail.com
	}%
	
	\date{\today}
	
	\begin{abstract}
		The ratios $\rho^{pp}_{\bar  pp}(s)$ of the real to the imaginary part of forward elastic $pp$ and $\bar pp$ scattering amplitudes at very high energies are considered in the models with rising total cross-sections and its difference. It is shown from the dispersion relations for  $pp$ and $\bar pp$ scattering amplitudes that in the Froissaron and Maximal Odderon approach  the ratios do not vanish asymptotically and they have the opposite signs for $pp$ and $\bar pp$ scattering.
	\end{abstract}
	
	\maketitle

\section*{Introduction}
It was claimed in the paper \cite{MW} that for elastic scattering amplitude the ratio $\rho=\text{Re}A(s,0)/\text{Im}A(s,0)$ has to be positive at $s\to  \infty$. However, this result can  not be considered as general one because of some assumptions are made indirectly.\footnote {In fairness, we note that in the Abstract of the journal version of the paper,  authors specify that the result relates to the crossing even component of the scattering amplitude} 
\begin{enumerate}
	\item Authors of the \cite{MW} have considered amplitudes for the processes $a+b \rightarrow a+b$ and $a+\bar b \rightarrow a+\bar b$ at high energies $\eta=\frac{1}{2}(s-u)=s-m_a^2-m_b^2 \gg m^2_{a,b}$. Then they write ''{\it Define $f(\eta)$ to be the average of the forward scattering amplitude for these two processes and $\sigma (\eta)$ that of the total cross sections, then $\text{Im}f(\eta)=\eta \sigma (\eta)$
	approximately at high energies.``}\footnote{To avoid any misunderstanding our notations below, the letter $\xi$ in the citation, which was used in \cite{AM}, is replaced for $\eta $} . 
	 It means that they consider only the crossing even part of the amplitude and their final conclusion about positive real part of the amplitude is true only for crossing even part of  the amplitude. 
    \item The special and  most interesting case we have if hadrons $a$ and $b$ are the protons. The crossing odd component, Odderon, for these amplitudes plays a very important role in observed differences in $pp$ and $\bar  pp$ cross  sections and it is lively discussed in the old and recent  papers \cite{Odd}.
    \item In order to make a conclusion about possible behavior of $\rho^{pp}_{\bar pp} (s)$ at $s\to \infty$ we should consider the most general case for odderon contribution allowed by the known restrictions on asymptotic properties of scattering $pp$ and $\bar pp$ amplitudes.

    This is the goal of the present Letter. Starting from the main strict results about crossing even and crossing odd $pp$ and $\bar pp$ amplitudes we will show  what we can say about total $pp$ and $\bar pp$ cross sections and $\rho^{pp}_{\bar pp}(s)$.
\end{enumerate}

\section*{Real part of the forward scattering amplitude}

 The crossing even and crossing odd amplitudes $A_+(s,t=0)=A_+(s), A_-(s,t=0)=A_-$ of the forward elastic $pp$ and $\bar pp$ scattering are defined as following
 \begin{equation}\label{eq:co-amps}
 f_{\pm}(s)=\dfrac{1}{2}\left [f^{pp}(s)\pm f_{\bar pp}(s)\right], \qquad f(s) \equiv  A(s,0)
 \end{equation}
 where  $m$ is the mass of proton.  Normalization of amplitudes is defined by the optic theorem in the following form
 \begin{equation}\label{eq:norm}
 \sigma_t(s)=\dfrac{\text{Im}A(s)}{s\sqrt{1-4m^2/s}}
 \end{equation}
We use here the following main facts concerning the amplitudes and cross sections under interest.
\begin{itemize}
\item [A.]Froissart-Martin-Lukazsuk bound \cite{FML}
\begin{equation}\label{eq:flm-bound}
 \sigma_t(s)\leq \dfrac{\pi }{m_{\pi }^2}\ln^2(s/s_0), \quad s_0=1\text{GeV}^2.
\end{equation}
In what follows we consider an arbitrary rise of cross section
\begin{equation}\label{eq:siggen}
\sigma_t(s)\propto \ln^{\alpha }(s/s_0), \quad 0<\alpha \leq 2.
\end{equation}
\item[B.]Bound on the difference of $pp$ and $\bar pp$ cross sections \cite{Eden}
\begin{equation}\label{eq:deltasig}
\begin{array}{ll}
\Delta \sigma_t&=|\sigma_t^{pp}(s)-\sigma_t^{\bar pp}(s)|\\&=\dfrac{2}{s\sqrt{1-4m^2/s}}|\text{Im}A_-(s,0)|
\propto \ln^\beta (s/s_0)
\end{array}
\end{equation}
where $\beta \leq \alpha/2$.
\item[C.]  The amplitudes $f_{\pm}(s)$ are analytic functions of $s$ in whole complex plane. These amplitudes satisfy the  twice subtracted dispersion relations because of $f_+(s)\propto s\ln^{\alpha }(s/4m^2)$ and $f_-(s)\propto s\ln^{\beta }(s/4m^2)$ at $s\gg 4m^2$, and $\alpha, \beta >0$.
\end{itemize}
\begin{equation}\label{eq:dispint+0}
\dfrac{\text{Re}f_{+}(s)}{s}=\dfrac{f_+(0)}{s}+\frac{2s}{\pi }{\cal P}\int\limits_{4m^2}^{\infty}\,
\frac{ds'}{(s'^{2}-s^{2})}\dfrac{\text{Im} f_{+}(s')}{s'}
\end{equation}
where ${\cal P}$ means principal integral value,
\begin{equation}\label{eq:dispint-0}
\dfrac{\text{Re}f_{-}(s)}{s}=f'_-(0)+\frac{2s^2}{\pi }{\cal P}\int\limits_{4m^2}^{\infty}\,
\frac{ds'}{(s'^{2}-s^{2})}\dfrac{\text{Im} f_{-}(s')}{s'^2}.
\end{equation}

Our aim is to find an asymptotic behavior of the real part of leading terms in crossing even and crossing odd amplitudes making use of Eqs. \eqref{eq:dispint+0} and \eqref{eq:dispint-0} taking a general form of crossing even and odd contributions \eqref{eq:siggen} and \eqref{eq:deltasig}. It would be sufficient in the case to use the derivative dispersion relations (DDR). They were suggested in \cite{BKS}. One can find more details  in the Refs. \cite{KN},   \cite{CMS} and \cite{AM}.

Let's consider the dispersion relations for arbitrary (but rising with energy) Pomeron and Odderon with the bounds \eqref{eq:siggen}, \eqref{eq:deltasig}. At $s \to \infty$ one can use the following approximation for $f_{\pm}(s)$
\begin{equation}\label{eq:pomodd}
\text{Im}f_{\pm}(s)/s\approx r_{\pm} \left \{
\begin{array}{ll}
&\xi^\alpha, \quad \alpha \leq 2,\\
&\xi^\beta, \quad \beta \leq \alpha/2,
\end{array}
\right. \quad \xi=\ln(s/4m^2).
\end{equation}

Making use of the Eq. \eqref{eq:pomodd} and the method to obtain DDR for $\text{Re}f_{\pm}(s)$ at $s\to \infty$, described in the Appendix,  we can write

\begin{equation}\label{ddr-as}
\dfrac{\text{Re} f_{\pm}(s)}{s}\approx r_{\pm}\left \{
\begin{array}{ll}
&\tan\left (\dfrac{\pi }{2}\hat{d}\right)\xi^\alpha=\dfrac{\pi }{2}\hat{d}(1+{\cal O}(\hat{d}^2))\xi^\alpha \\ &\approx \dfrac{\pi }{2}\alpha \xi^{\alpha-1},\\
&-\cot\left (\dfrac{\pi }{2}\hat{d}\right)\xi^\alpha=-\frac{1-\frac{1}{2} \left (\frac{\pi \hat{d}}{2}\right )^2+ \dots }{\frac{\pi \hat{d}}{2}-\frac{1}{3}\left (\frac{\pi  \hat{d}}{2}\right )^3+ \cdots }\xi^{\beta }\\
&\approx -\dfrac{2}{\pi }\hat{d}^{-1}\xi^{\beta }(1+{\cal O}(\hat{d}^2))\\ &=-\dfrac{2}{\pi }\int d\xi \xi^\beta \approx -\dfrac{2}{\pi }\dfrac{1}{1+\beta }\xi^{\beta+1}
\end{array}
\right .
\end{equation}
where $\hat d=d/d\xi$. The sign ''-`` in $\text{Re}f_-(s)$  is originated from our definition of the amplitudes $f_\pm$ in Eq. \eqref{eq:co-amps}.

For the leading terms at $s\to \infty$ we have
\begin{equation}\label{eq:fplus}
\dfrac{1}{s}\text{Re}f_+(s)=r_+\dfrac{\pi }{2}\alpha \xi^{\alpha-1},
\end{equation}
\begin{equation}\label{eq:fminus}
\dfrac{1}{s}\text{Re}f_-(s)=-r_-\dfrac{2}{\pi (1+\beta )} \xi^{1+\beta },
\end{equation}
\begin{equation}\label{eq:fpp-barpp}
\text{Re}f^{pp}_{\bar pp}(s)=s\left [\dfrac{\pi }{2}r_+\alpha \xi^{\alpha-1}\mp r_-\dfrac{2}{\pi (1+\beta )} \xi^{1+\beta }\right].
\end{equation}

If we consider parameters $\alpha$ and $\beta$ in region 
\begin{equation}\label{eq:alpha-beta}
0<\alpha\leq 2  \qquad 0<\beta\leq \alpha/2
\end{equation}
which corresponds to the models with infinitely rising $\sigma_{t}$ and $|\Delta \sigma_{t} |$
we find that the second term in Eq. \eqref{eq:fpp-barpp} dominates at $\xi\to \infty$ because of $\beta+1> 1$ and $\alpha-1\leq 1$.

So, {\bf the first conclusion} is the following. 

The real part of the $pp$ and $\bar pp$ scattering amplitude in the models with infinitely rising cross sections and difference of the cross sections is asymptotically dominated by Odderon contribution.

\begin{figure}[!h]
	\centering
	\includegraphics[width=0.9\linewidth]{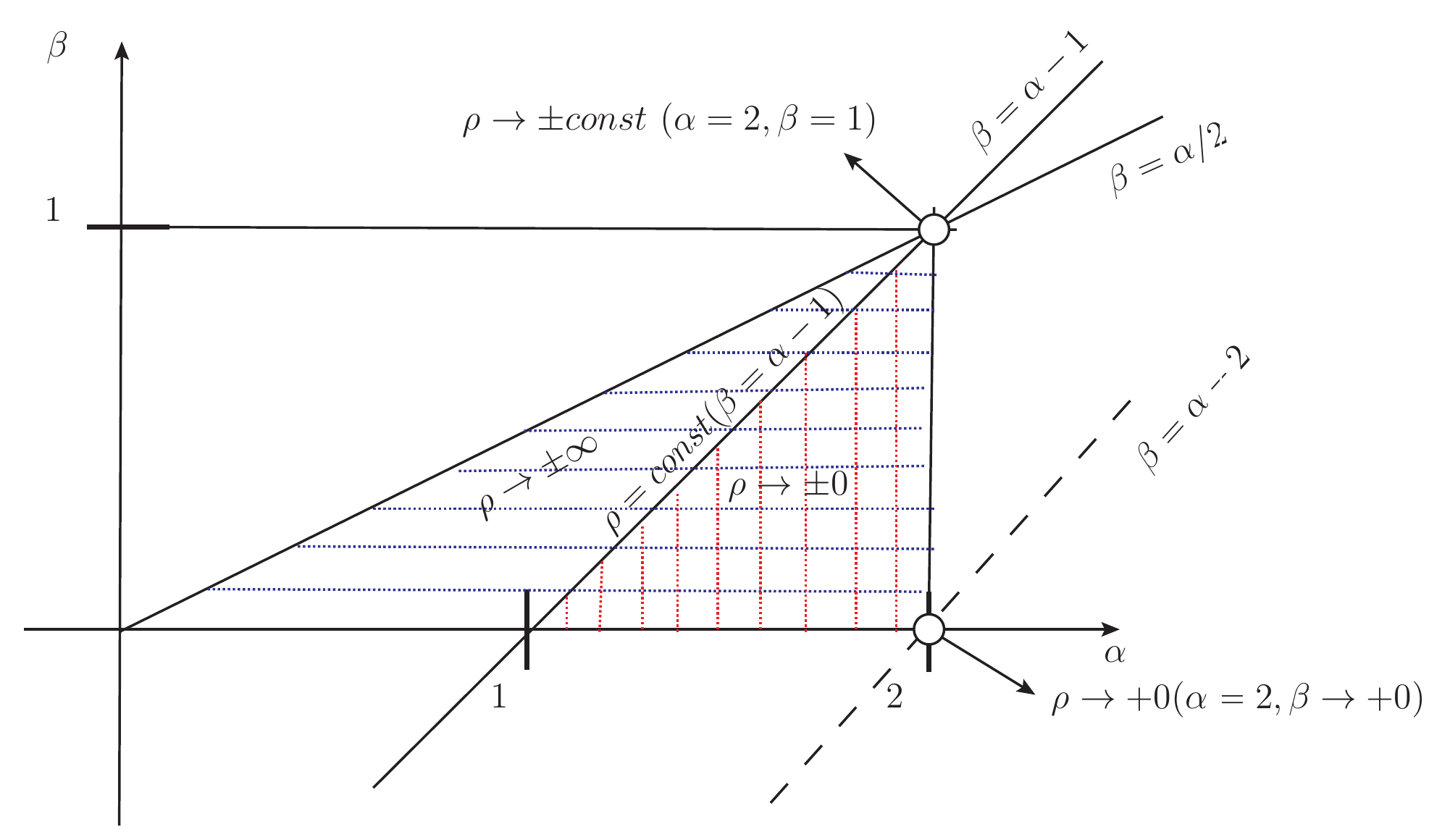}
	\caption{Ratio $\rho^{pp}_{\bar pp}$ at $s\to \infty$ in different regions of the plane ($\alpha,\beta$)}
\label{fig:rho-fmo}
\end{figure}

Thus, in this case
\begin{equation}\label{eq:rho@rising-sigma}
\rho^{pp}_{\bar pp}(s)=\dfrac{\text{Re}f^{pp}_{\bar pp}(s)}{\text{Im}f^{pp}_{\bar pp}(s)}\approx \mp \dfrac{r_-}{r_+}\dfrac{2/\pi}{1+\beta}\xi^{1+\beta-\alpha}.
\end{equation}
If $0<\beta<\alpha-1$ then at $\xi\to \infty$ ratios $\rho(\xi)\to \pm 0$ (with the opposite signs for $pp$ and $\bar pp$).
 But if $\alpha-1<\beta<\alpha/2$ then ratios $\rho(\xi)\to \pm \infty$.  
This is shown on Fig. \ref{fig:rho-fmo}. 

In the case of the Maximal Odderon at fixed $0<\alpha \leq 2$ we have $\beta =\alpha/2$.
Then at any allowed positive $\alpha$
\begin{equation}\label{eq:maxodd}
\begin{array}{ll}
\text{Re}f^{pp}_{\bar pp}(s)&=s\left [\dfrac{\pi }{2}r_+\alpha \xi^{\alpha-1}\mp r_-\dfrac{2}{\pi (1+\alpha/2)} \xi^{1+\alpha/2}\right ]\\
&\approx \mp s\dfrac{2r_-}{\pi (1+\alpha/2)}\xi^{1+\alpha/2}.
\end{array}
\end{equation}
Consequently
\begin{equation}\label{eq:rhomax}
\rho^{pp}_{\bar pp}(s)=\dfrac{\text{Re}f^{pp}_{\bar pp}(s)}{\text{Im}f^{pp}_{\bar pp}(s)}\approx \mp \dfrac{2r_-/\pi }{r_+}\dfrac{1}{1+\alpha/2}\xi^{1-\alpha/2}.
\end{equation} 
We would like to notice (it is {\bf the second conclusion}) that $|\rho^{pp}_{\bar pp}(s)|\to \infty$ at $s\to \infty$ if $0<\alpha <2$.  

As well we have {\bf the third conclusion}: $|\rho^{pp}_{\bar pp}(s)|\to \text{const} \neq  0$ if $r_- \neq  0$ only in the case of the Froissaron and Maximal Odderon ($\alpha=2, \beta =1$).

In the FMO model \cite{MN1,MN2} we have considered and compared with the latest data of TOTEM \cite{TOTEM1, TOTEM2} the case $\alpha=2, \beta=1$. Performing fit with arbitrary values of $\alpha$ and $\beta$ \cite{MN2} we have found that $\alpha, \beta$ come back to the maximal values 2,1, correspondingly.

In \cite{MN1} the leading terms were parameterized in the form
\begin{equation}\label{eq:fmo-param}
\dfrac{k}{s}f_{\pm}=\left \{
\begin{array}{ll}
&iH_1\tilde \xi^2\approx iH_1\xi^2+\pi H_1\xi,\\
&O_1\tilde \xi^2\approx O_1\xi^2-iO_1\pi \xi.
\end{array} 
\right .
\end{equation}
where $k=0.3894...\,\,  \text{mbGeV}^2$ and $\tilde \xi=\xi- i\pi/2$ .

Comparing the Eqs. \eqref{eq:fplus}, \eqref{eq:fminus} with \eqref{eq:fmo-param} we have $r_+=H_1/k, \quad r_-=-\pi O_1/k$ where $H_1=0.25 \text{mb}, O_1=-0.05 \text{mb}$ \cite{MN1} .
Thus
\begin{equation}
\lim_{s\to \infty}\rho^{pp}_{\bar pp}(s)=\mp\dfrac {r_-}{r_+\pi } = \pm\dfrac {O_1}{H_1}=\mp 0.2
\end{equation}

\begin{figure}[!h]
	\centering
	\includegraphics[width=1.\linewidth]{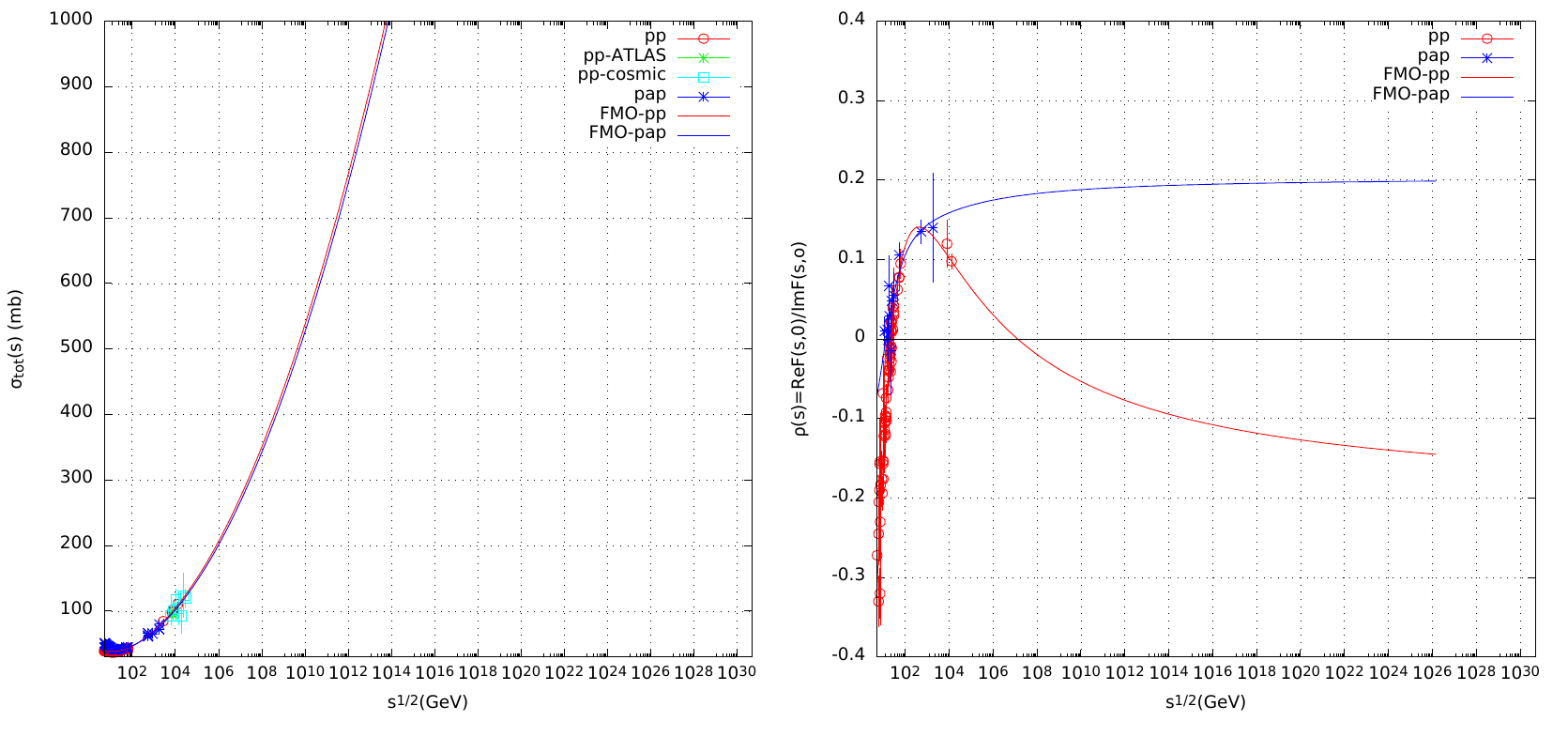}
	\caption{Extrapolation of $\sigma_t(s)$ and $\rho (s)$ in the FMO model at $t=0$ \cite{MN1}}
	\label{fig:fmo-superhigh-s}
\end{figure}

At the Fig. \ref{fig:fmo-superhigh-s} we show an extrapolation of the result obtained in \cite{MN1} $\rho_{\bar pp}^{pp}$ for higher energies. One can see that the real asymptotic regime occurs at extremely high energy. Even a change of the sign in $\rho^{pp}(s)$ is attained at $\sqrt{s} \sim 10^4$ TeV.

\section*{Conclusion}

We have shown that the ratios $\rho^{pp}_{\bar  pp}(s)$ of the real to imaginary part of forward elastic $pp$ and $\bar pp$ scattering is not positive for the both $pp\to pp$ and $\bar pp\to \bar pp$ processes with an odderon contribution to the amplitudes do not vanish at $s\to \infty$.  

We would like to emphasize that such a regime is confirmed by comparison of the Froissaron and Maximal Odderon approach \cite{MN1,MN2} with the experimental data on forward $pp$ and $\bar pp$ scattering including the latest TOTEM data.  The model predicts asymptotic values of the ratios $\rho^{pp}_{\bar  pp}(s\to \infty)\approx \mp 0.2$.

\begin{acknowledgments} The authors thank Prof. Basarab Nicolescu for a careful reading of the manuscript and interesting, fruitful discussions. E.M. thanks the Department of Nuclear Physics and Power Engineering of the National Academy of Sciences of Ukraine for support (the project No 0118U005343). 
\end{acknowledgments}

\section*{Appendix}\label{sec:appendix}
The integrals in \eqref{eq:dispint+0} and \eqref{eq:dispint-0} (without unimportant at $s\to \infty$ subtracted terms) can be presented in the common form with $\nu=1$ for $\text{Im}f_+(s)$ and $\nu=0$ for $\text{Im}f_-(s)$
\begin{equation}\label{eq:dispint+}
\begin{array}{rl}
\dfrac{\text{Re}f(s)}{s}&=\dfrac{2}{\pi }s^{2-\nu }{\rm P}\int\limits_{4m^2}^{\infty}\,
 \dfrac{ds'}{s'^{2}-s^{2}} g(s')\\
&=\dfrac{2}{\pi }e^{(2-\nu )\xi }{\rm P}\int\limits_{0}^{\infty}\,
\dfrac{e^{\xi'}d\xi'}{e^{2\xi'}-e^{2\xi }} g(\xi')
\end{array}
\end{equation}
where
$$g(s')=(s')^{\nu-1}\text{Im}f(s')/s'$$ or  $$g(\xi')=e^{(\nu-1)\xi'}e^{-\xi'}\text{Im}f(\xi').$$
After some simple transformation one can obtain
\begin{equation}\label{eq:dint-1}
\begin{array}{ll}
e^{-\xi }\text{Re}f(\xi )&=\dfrac{1}{\pi }e^{(1-\nu )\xi }\left \{\ln\dfrac{e^{\xi }+1}{e^{\xi }-1}g(\xi'=0)\right .\\
&\left .+ \int\limits_0^\infty d\xi'\, \ln \dfrac{ e^{\xi }+e^{\xi^\prime}}{|e^\xi-e^{\xi'}|}
 \left (e^{(\nu-1)\xi'}g(\xi')\right )'\right \}
\end{array}
\end{equation}
The logarithmic factor in the integral \eqref{eq:dint-1} can be transformed as follows
$$
\begin{array}{rl}
\ln\dfrac{e^{\xi'}+e^{\xi }}{|e^{\xi'}-e^{\xi }|}&=
\ln\dfrac{e^{(\xi'-\xi )/2}+e^{-(\xi'-\xi )/2}}{|e^{(\xi'-\xi )/2}-e^{-(\xi'-\xi )/2}|}\\
&=\ln|\coth\dfrac{1}{2}(\xi'-\xi )|
=\ln\dfrac{1+e^{-|x|}}{1-e^{-|x|}}\\
&=2\sum\limits_{p=0}^{\infty}\dfrac{e^{-(2p+1)|x|}}{2p+1}, \quad x=\xi'-\xi.
\end{array}
$$

All other factors in the \eqref{eq:dint-1} can be expanded in powers of $\xi'-\xi$
\begin{equation}\label{eq:expansions}
\begin{array}{l}
\left (e^{(\nu-1)\xi'}g(\xi')\right )'=e^{(\nu-1)\xi'}\left (\nu-1+\dfrac{d}{d\xi'}\right )g(\xi'),\\
\tilde g(\xi')=\left (\nu-1+\dfrac{d}{d\xi'}\right )g(\xi')=\sum\limits_{k=0}^{\infty}\frac{(\xi'-\xi )^{k}}{k!}\hat{d}^{k}\tilde
g(\xi ),\\
e^{(1-\nu )\xi }e^{(\nu-1)\xi'}=
\sum\limits_{n=0}^{\infty}\dfrac{(\nu-1)^{n}}{n!}(\xi'-\xi )^{n}.\\
\hat d=d/d\xi .
\end{array}
\end{equation}

Taking into account the above expression and omitting the first term in \eqref{eq:dint-1} because it goes to 0 at $\xi \to \infty$ one can write the
integral  \eqref{eq:dint-1} in the form
\begin{equation}\label{eq:dispint1}
\begin{array}{ll}
e^{-\xi }\text{Re}f(\xi )&=
\dfrac{2}{\pi }\sum\limits_{p=0}^{\infty}\dfrac{1}{2p+1}
\sum\limits_{k=0}^{\infty}\sum\limits_{n=0}^{\infty}\dfrac{(\nu-1)^{n}}{k!n!}\\
& \times  {\cal I}(\xi;p,k,n) \cdot \hat d^{k}\tilde g(\xi )
\end{array}
\end{equation}
where
$$
\begin{array}{ll}
{\cal
	I}(\xi;p,k,n)&=\int\limits_{0}^{\infty}d\xi'e^{-(2p+1)|\xi'-\xi|}(\xi'-\xi )^{k+n}\\
&=(-1)^{k+n}\int\limits_{0}^{\xi }d\xi'e^{-(2p+1)(\xi-\xi')}(\xi-\xi')^{k+n}\\
&+
\int\limits_{\xi }^{\infty}d\xi'e^{-(2p+1)(\xi'-\xi )}(\xi'-\xi )^{k+n}\\
&=\dfrac{1}{(2p+1)^{k+n+1}}\left [ \Gamma (k+n+1) )\right .\\
&\left . +
(-1)^{k+n}\gamma (k+n+1,\xi (2p+1)\right ] \nonumber
\end{array}
$$
and $\gamma (a,x)$ is incomplete gamma function. 

At fixed $a$ and $x\to \infty$
\begin{equation}\label{eq:gamma}
\gamma (a,x)=\Gamma (a)-e^{-x}x^{a}(1+{\cal O}(1/{x}),
\end{equation}
therefore we have
\begin{equation}
\begin{array}{ll} 
e^{-\xi }\text{Re}f(\xi )&\approx 
\dfrac{2}{\pi }\sum\limits_{p=0}^{\infty}\dfrac{1}{(2p+1)^2} 
\sum\limits_{k=0}^{\infty}\sum\limits_{n=0}^{\infty}\dfrac{(k+n)!}{k!n!}
\\
& \times \left (\dfrac{\nu-1}{2p+1}\right )^{n} \left (\dfrac{\hat
	d}{2p+1}\right )^{k}(1+(-1)^{k+n})\tilde g(\xi )\\
&=
\dfrac{2}{\pi }\sum\limits_{p=0}^{\infty}\dfrac{1}{(2p+1)^2} 
\left \{\dfrac{1}{1-(\nu-1+\hat
	d)/(2p+1)}\right .\\
&\left . +\dfrac{1}{1+(\nu-1+\hat d)/(2p+1)} \right \}\tilde g(\xi )\\
&= \dfrac{4}{\pi }\sum\limits_{p=0}^{\infty}
\dfrac{1}{(2p+1)^{2}-(\nu-1+\hat d)^{2}}\tilde g(\xi ).\\ \nonumber
\end{array}
\end{equation}  
The last sum in the above expression is simplified making use of the  (see \cite{BMP})  
\begin{equation}
\sum\limits_{k=0}^{\infty}\frac{1}{(2p+1)^{2}-a^{2}}=\frac{\pi }{4a}\tan(\pi
a/2), \nonumber
\end{equation}

Thus, we have finally the asymptotic form of the derivative dispersion relations
\begin{equation}\label{ddr+}
\Re ef_{+}(s)/s\approx \tan \left ( \frac{\pi }{2}
\hat{d}\right )\text{Im}f_{+}(s)/s, 
\end{equation}
\begin{equation}\label{ddr-}
\begin{array}{ll}
\text{Re}f_{-}(s)/s&\approx \tan \left [ \frac{\pi }{2}(-1+\hat{d})\right ]\text{Im}f_{-}(s)/s\\
&=- \cot \left ( \frac{\pi }{2}\hat{d}\right )\text{Im}f_{-}(s)/s. 
\end{array}
\end{equation}

\hspace{2.cm}

\end{document}